\documentclass{article}

\usepackage{arxiv}
\usepackage{cite}

\usepackage[utf8]{inputenc} 
\usepackage[T1]{fontenc}    
\usepackage{hyperref}       
\usepackage{url}            
\usepackage{booktabs}       
\usepackage{amsfonts}       
\usepackage{nicefrac}       
\usepackage{microtype}      
\usepackage{lipsum}
\def\BibTeX{{\rm B\kern-.05em{\sc i\kern-.025em b}\kern-.08em
    T\kern-.1667em\lower.7ex\hbox{E}\kern-.125emX}}
\usepackage{comment}
\usepackage{multirow}
\usepackage{float}

\title{An Exploratory Study of How Specialists Deal with Testing in Data Stream Processing Applications}

\author{
  Alexandre Vianna\thanks{This research was funded by INES 2.0, grants FACEPE PRONEX APQ 0388-1.03/14 and CNPq 465614/2014-0.} \\
  Centro de Inform\'atica - CIn\\
  Federal University of Pernambuco\\
  Recife, Brazil \\
  \texttt{asgv@cin.ufpe.br} \\
   \And
    Waldemar Ferreira\\
  Centro de Inform\'atica - CIn\\
  Federal University of Pernambuco\\
  Recife, Brazil \\
  \texttt{wpfn@cin.ufpe.br} \\
  \And
    Kiev Gama\\
  Centro de Inform\'atica - CIn\\
  Federal University of Pernambuco\\
  Recife, Brazil \\
  \texttt{kiev@cin.ufpe.br} \\
}

\begin{document}
\maketitle

\begin{abstract}
[Background] Nowadays, there is a massive growth of data volume and speed in many types of systems. It introduces new needs for infrastructure and applications that have to handle streams of data with low latency and high throughput. Testing applications that process such data streams has become a significant challenge for engineers. Companies are adopting different approaches to dealing with this issue. Some have developed their own solutions for testing, while others have adopted a combination of existing testing techniques. There is no consensus about how or in which contexts such solutions can be implemented. [Aims] To the best of our knowledge, there is no consolidated literature on that topic. The present paper is an attempt to fill this gap by conducting an exploratory study with practitioners. [Method] We used qualitative methods in this research, in particular interviews and survey. We interviewed 12 professionals who work in projects related to data streams, and also administered a questionnaire with other 105 professionals. The interviews went through a transcription and coding process, and the questionnaires were analysed to reinforce findings. [Results] This study presents current practices around software testing in data stream processing applications. These practices involve methodologies, techniques, and tools. [Conclusions] Our main contribution is a compendium of alternatives for many of the challenges that arise when testing streaming applications from a state-of-the-practice perspective.
\end{abstract}


\section{Introduction}

In recent years, there has been a massive growth in the amount of data generated by mobile phones, IoT devices, and user interaction on social networks, websites, and applications. Easily, billions of events can be triggered per minute. Processing this data is a challenge. Techniques in the realm of Big Data were proposed to overcome this issue~\cite{chen2014data, kaisler2013big}. For those applications that require low latency, data streams have emerged as a branch of Big Data to handle real-time applications~\cite{liu2014survey}. Data Stream Processing (DSP) approaches are recommended for real-time data processing. It applies specific techniques for capturing and processing relevant data on-the-fly, i.e., without requiring storage~\cite{stonebraker20058}.

One of the main challenges when developing DSP applications is testing them. Like many other software areas, testing plays an essential role in the quality assurance of DSP applications. However, testing such kinds of software is not an easy task. Some factors that hinder testing are message temporality, parallelism, data volume, variability, speed of messages, among others.  Developing efficient and effective streaming tests is complex. So, the importance of testing is apparent in many sources such as tools discussions lists, talks at technical events, professional social networks, and open-source code repositories.

Concerning DSP applications, the industry know-how is vast but is neither uniform nor widely disseminated or consolidated in the community. Furthermore, there is limited literature about testing DSP applications. Besides, there is a gap between industrial and academic knowledge about the testing of data stream applications. To the best of our knowledge, there was no research synthesizing the experience of practitioners and summarising the state-of-the-practice about testing DSP .

In this article, we performed an exploratory study to find and report relevant industry knowledge around software testing in DSP applications. We carried out 12 interviews with practitioners and administered a questionnaire with 105 professionals, in which among those we found employees of companies figuring in top rankings from Fortune~\cite{Fortune} and Forbes~\cite{Forbes}.


\section{Background}
\label{sec:back}
\subsection{Data Stream Processing}


A DSP approach can be understood as a sequence of continuously generated data, in which the order of data arrival cannot be controlled~\cite{stephens1997survey}. These applications do not store any data, with all data being computed in real-time. DSPs use sliding windows in which patterns have to be detected with very low latency~\cite{babcock2002models}. To achieve real-time processing, it often involves distributed computing systems~\cite{cherniack2003scalable}.

This DSP approach comprises filters based on a sequence of operation-based transformations defined by relational algebra~\cite{cugolaf2012processing}. The industry has adopted such an approach for applications with high data volume that require real-time answers. For example, anti-fraud systems, real-time dashboards, clickstream for user behaviour analysis~\cite{hanamanthrao2017real}, analysis of user engagement in marketing campaigns~\cite{hasan2018survey}, geofences, investment robots, and equipment failure detection.


\subsection{Related Work}

Although there is no literature overview of testing practices in DSP applications, there are works focusing on specific tools and approaches and some studies on topics related to testing and quality of data stream software.

Zvara et al.~\cite{zvara2017tracing} presented a tracing framework designed for streaming systems. It can be used to debug DSP applications through tracing of individual input records. This tool helps to identify common problems such as bottlenecks, irregular characteristics of incoming data, anomalies propagated unexpectedly through the system in real-time.  This approach is valuable to avoid developers from spending countless hours on identifying such problems manually.

Restream~\cite{schleier2016restream} is a tool that replays streams from historical event logs. This tool is designed for accelerated replay, preserving sequential semantics. Although ReStream simulates streams precisely in a controlled environment, there is a lack of simulation of fault tolerance scenarios. On the contrary, it handles fault tolerance cases by avoiding them during the simulation, such as ignoring duplicated messages.  

Due to the large volumes of data, a good test data is relevant when testing data stream software. When real historical data is not available, test generation can be an option. Gutierrez et al. proposed a tool~\cite{gutierrez2018iot} that make use of high-level languages for describing events patterns and rules for event test generation.  It has controlled randomization to generate values within a set between a minimum and maximum value. 


\section{Research Design}
\label{sec:method}

We conducted an exploratory study inspired by qualitative methods for software engineering~\cite{seaman1999qualitative}. We carried interviews and administered questionnaires to a group of practitioners. The primary technique used in data analysis was coding~\cite{saldana2015coding}. Moreover, we double-checked our findings by comparing our results from the interviews and questionnaires.

We can summarise our goal with this work as an inquiry about how the industry is testing DSP application. This goal leads us to the following research questions:

\textbf{RQ1:} \textit{ What are the approaches and techniques being adopted to test data stream processing applications? }

\textbf{RQ2:} \textit{ Which test frameworks and tools have been adopted?}

\textbf{RQ3:} \textit{ What kind of data is being used to test data stream processing applications?}

\subsection{Data Collection Techniques}

To increase our confidence in our results, we carried out interviews and an on-line survey.

\subsubsection{Interviews}
\label{sec:inter}

We conducted semi-structured interviews with 12 practitioners. We selected volunteers with at least one year of professional experience in developing DSP projects. The interviewees were contacted through the university's contact network and manual searches of curricula in professional social networks. The participants signed an informed consent term agreeing to the interview's recordings and anonymity. The interviews were conducted over Skype calls ranging from 41 to 54 minutes. 

The interview script is available in~\cite{PaperAttachments}. Each interview had three sections. The first part explored the interviewee's background, experience, current activities, and company profile. The second part has technical questions about DSP projects and how the company deals with data streams (tools, mechanisms, processes, etc.). Finally, the third part concentrated in knowing the companies process on testing DSP applications. In particular, we focused on asking about tools and standardised methodologies. 

\subsubsection{Questionnaire}
\label{sec:quest}

We also administered questionnaires with statements about data stream, aiming at a broader audience. The full questionnaire is available~\cite{PaperAttachments}. The questionnaires were disseminated in forums of stream processing tools and direct contact via professional social networks. By the end of three months, we collected 105 responses, but only 101 responses were considered due to inconsistencies found in 4 responses. A sample of 101 responses is satisfactory when compared to other online surveys in software engineering~\cite{punter2003conducting}.

The questionnaire was formulated by the first author, and the questions were based on the experience gathered from the first two interviews. The questionnaire followed a script similar to that of the interviews, being slightly adjusted after comments from other authors and validated by a DSP expert.

\subsection{Data Analysis Techniques}

The main technique used to analyse data collected was coding~\cite{saldana2015coding}. As recommended by Glaser et al.~\cite{glaser1992basics}, we initially read and systematically coded all interview transcripts. It created an initial set of codes. After we clustered this initial set of codes in categories by comparing new incidents with previous codes, several categories emerged during this phase. We then compared categories from one interview with categories from other interviews, delimiting coding to only those categories that are related to the core category -- \textbf{Testing Data Stream Software}. After that, we matched our findings (categories and relationships) with answers from open-ended questions of the questionnaires. We coded those passages that confirm or contrast with any findings from interviews. This data triangulation aims to reduce researcher bias. Using only one data source is a threat because sometimes what people report in an interview and questionnaire is not always consistent. A classic example is that people tend to underestimate how long it will take to complete tasks~\cite{kruger2004if}.

Finally, the answers for each research question was based on the categories and codes that emerged from the interviews and questionnaires. Our last step was to provide conceptual relations between research questions answers and the existing literature on the data streaming field.

\section{Demographics}
\label{sec:demo}

Throughout the text, interviewees will be referenced as P[1-12] according to Table~\ref{tab:profile}, while the questionnaire respondents will be mentioned as Q[1-101].

\noindent\textbf{Interviews.} We ended up reaching 12 participants with different profiles in both experience and profile of the companies and projects in which they work. Table~\ref{tab:profile} summarizes the participants' profile.  Although some of them have short professional experience time, they have enough to contribute by being involved in companies with relevant cases in the data stream area (e.g., participant P9 has one year of experience but deals with over 400,000 requests/day). We also interviewed very experienced professionals (such as P2, P6, and P10) with more than 15 years of working in the area. Participant P6 works with DSP in a large bank and has a broad background in the Big Data domain, having certifications in stream processing platforms, besides having worked in several companies in the financial sector. The level of education is high (P1, P2, and P8 have MSc degree, and P10 a Ph.D. degree). Company profiles are diverse, ranging from start-ups with few professionals, who  developed innovative services involving DSP, to traditional large companies (some are listed on top rankings from Fortune~\cite{Fortune} and Forbes~\cite{Forbes}) with thousands of IT employees, using DSP solutions in their operations. This diverse spectrum allows us to capture different experiences and views on the subject.

\begin{table*}[]
\centering
\caption{Interview Participant Profile. ITX means IT experience in years, DSX means data stream development experience in years, ITE means number of IT employees in the company, and DSP the estimated number of Data Stream Processing developers. }
\begin{tabular}{lllrrlrr}
\hline
\textbf{P\#} & \textbf{\begin{tabular}[x]{@{}l@{}}Education\\Level\end{tabular}} & \textbf{Job Title}                                             & \textbf{ITX} & \textbf{DSX} & \textbf{Company Main Sector}                                                         & \textbf{\#ITE}	&  \textbf{\begin{tabular}[x]{@{}l@{}}DSP\\Developers\end{tabular} } \\ \hline
P1		&Master           & Backend Engineer     & 14           & 2            & \begin{tabular}[x]{@{}l@{}}Financial (Benefits \&\\Rewards Services)\end{tabular} & 8+	& 2             \\ \hline
P2		&Master           & Software   Architect & 21           & 4            & Software Consulting           & 2000+	& 7          \\ \hline
P3		&Bachelor           & Software Engineer    & 8            & 2            & Location Data Services       & 70+	& 6            \\ \hline
P4		&Bachelor           & Software   Engineer  & 9            & 3            & \begin{tabular}[x]{@{}l@{}}Financial (Banking\\Services on App)\end{tabular} & 200+	& 100           \\ \hline
P5		&Bachelor           & Software Engineer    & 3            & 2            & \begin{tabular}[x]{@{}l@{}}Financial (Banking\\Services on App)\end{tabular}        & 200+	& 100           \\ \hline
P6		&Bachelor           & Big Data   Engineer  & 16           & 4            & \begin{tabular}[x]{@{}l@{}}Financial\\(Traditional Bank)\end{tabular}             & 5000+	& -          \\ \hline
P7		&\begin{tabular}[x]{@{}l@{}}Master\\(in progress)\end{tabular}            & Data Engineer    & 10           & 4            & \begin{tabular}[x]{@{}l@{}}Financial\\(Investment Broker)\end{tabular}   & 1600+	& 8          \\ \hline
P8		&Master           & Software   Engineer  & 6            & 3            & Software  Consulting         & 2000+	& 100          \\ \hline
P9		&Bachelor           & Data Engineer        & 1            & 1            & Food Delivery Service         & 250+	& 50           \\ \hline
P10		&PhD          & Software   Engineer  & 22           & 10           & Solutions for   Smart Cities  & 10+	& 5            \\ \hline
P11		&Bachelor          & Data Engineer        & 3            & 2            & Data Science Solutions         & 11+	& 3            \\ \hline
P12		&Bachelor          & Software  Developer & 8            & 4            & E-commerce   & 40+	& 6            \\ \hline
\end{tabular}
\label{tab:profile}
\vspace{-6mm}%
\end{table*}

\noindent\textbf{Questionnaire.}
The questionnaire participants profile is quite varied in terms of experience and company business sector. They have a high degree of qualification, where 68 out of the 101 participants have a master's or doctorate degrees. 
An interesting finding is the significant number of participants associated with research and development (R\&D) of tools for stream processing. This number may be associated with the dissemination of the questionnaires that were also made through discussion forums in DSP tools communities. Another interesting finding is that most of the respondents with R\&D were in industry instead of academia. As we can see, our sample includes developers of data stream tools with a high level of training, which highlights a relevant dimension of this work.

\section{Categories and Concepts}
\label{sec:cat}

As mentioned in Section~\ref{sec:method}, we analysed our data following some coding techniques to answer our research questions. With these techniques, the core category  \textit{Testing Data Stream Software} emerged. To answer each research question, three subcategories emerged: \textit{Testing Approaches}, \textit{Testing Tools} and \textit{Test Data}. In the next sections, we present each category.

\subsection{The Core Category: Testing Data Stream Software}

Testing emerged as a highlighted issue for data stream applications, with more than one hundred code references. For instance, a comment from interviewee P6 reveals the importance of quality assurance in a DSP of anti-fraud features in the financial sector. This is crucial because failures result in substantial losses in a short time.

\begin{center}\noindent\fbox{%
    \parbox{0.6\columnwidth}{%
        \textit{``The anti-fraud feature processes the user interaction data stream. It checks in real-time against a previously trained model if these data fit the pattern of regular interaction or a fraudulent interaction. What is the cost of false negatives? It can be huge."} (P6, Data Engineer, Financial)
    }%
}
\end{center}

In some interviews, participants commented on specific issues that threat DSP application testing. The most frequent problems are message synchronization, multiple data streams operations, data heterogeneity, fault tolerance, difficulties on traceability, and generating test data. Below, a response from Q-59 highlights message synchronization issues.

\begin{center}\noindent\fbox{%
    \parbox{0.6\columnwidth}
    {%
\textit{``Our data stream is immensely partitioned and duplicated. We have a complicated tumbling window with custom `go-next' logic requiring various events. Stream synchronization and window tumbling is the biggest challenge."} (Q-59)
    }%
}
\end{center}

Participants reported relevant aspects that hamper the development and testing of DSP applications. The quotation below synthesises these comments:

\begin{center}\noindent\fbox{%
    \parbox{0.6\columnwidth}{%
\textit{``The biggest challenge is scale. The problem is the data growth rate, so we reached a petabyte here, and it keeps growing. Then things start to get rough.''} (P9, Data Engineer, Food Delivery Service)
    }
}
\end{center}

Finally, the importance of testing was broadly commented both in interviews and questionnaires. Specific difficulties were also identified for testing data stream applications. Therefore, data stream testing emerges as a fruitful topic to explore as the core category of this work.

\subsection{ Category: Testing Approaches}
\label{sec:approaches}

The following subsections present a summary of the most recurring approaches on testing DSP applications.

\subsubsection{ Category: Levels of Testing} Traditional testing methodologies continue to be employed to test DSP applications. However, participants also reported adaptations and special configurations in traditional test levels for DSP applications. The comment of interviewee P6 illustrates this position:

\begin{center}\noindent\fbox{%
    \parbox{0.6\columnwidth}{%
\textit{``Software development has not changed, so unit testing and integration testing are continuing, but now considering data stream particularities."} (P6, Data Engineer, Financial)
    }
}
\end{center}

Regarding unit tests, 10 interviewees and 59 survey respondents indicate that unit tests are consolidated for testing data stream modules. In general, no particular difficulties were reported when implementing unit tests in the context of DSP. The comment of P7 is an example of this position:

\begin{center}\noindent\fbox{%
    \parbox{0.6\columnwidth}{%
\textit{``Usually, we test based on unit testing in development, and only after passing in all unit tests, we can deploy it to production."} (P7, Big Data Engineer, Financial)
    }
    
}
\end{center}

As for the integration test, no challenge has been reported, and it can be easily automated using test data mocking tools. However, difficulties start in system tests, because it involves all services running. Some participants reported building a copy of the production environment for system testing, but this requires an abundant infrastructure, and it is a costly solution. Comments of P2 reinforce challenges on system test:

\begin{center}\noindent\fbox{%
    \parbox{0.6\columnwidth}{%
\textit{``Here we perform system tests. But in the stream part, it is not that easy to test end-to-end. It involves raising and connecting several micro-services."} (P2, Software Architect, Software Consulting Sector)
    }%
}
\end{center}

Another issue highlighted by the participants was the relevance of load tests in system test levels to find typical data stream problems that occurs with system stress:

\begin{center}\noindent\fbox{%
    \parbox{0.6\columnwidth}{%
\textit{``Data stream has exacerbated the importance of load testing, especially in the stage environment with all modules installed."} (P6, Data Engineer, Financial)
    }
}
\end{center}

Finally, the following quotation synthesizes the testing process from the company of respondent Q-24. This comment corroborates with previous codes and observations: 

\begin{center}\noindent\fbox{%
    \parbox{0.6\columnwidth}{%
\textit{``i) Unit testing, locally on devs' machines ii) Integration testing, locally on devs' machines. We leverage Flink's ability to run stream-processing in a local environment. iii) End-to-end testing on staging environment. We use production test data from an in-house built generator."} (Q-24)
    }
}
\end{center}

\subsubsection{Stream Simulation}

Stream simulation is an approach to testing stream applications realistically. It involves replaying historical or generated data. Nineteen questionnaire participants reported performing stream simulation as a test approach.  
Simulation tools may provide features for handling typical problems of DSP applications, such as the message entry semantics, loss of messages, duplicate messages, corrupt messages, message asynchronicity, latency, hardware failures, throughput and others. In fact 83\% of respondents consider fault tolerance features helpful in data stream simulation tools. The following excerpt illustrates these comments:

\begin{center}\noindent\fbox{%
    \parbox{0.6\columnwidth}{%
\textit{``Simulating fault tolerance is interesting because with a history replay we can perform the processing and then assert the results. Meanwhile, there may be some scenarios of service dying and rising, machines with problems and latency on the network. We can check if even with this high entropy, the answers are consistent. "}(P4, Software Engineer, Financial)
    }
}
\end{center}

\subsection{Category: Testing Tools}
\label{sec:cattools}

Below we present test tools along with observations about the purpose of the tool. For unit tests there is the Spark Testing Base, which emerged in questionnaires and on P9's interview: 

\begin{center}\noindent\fbox{%
    \parbox{0.6\columnwidth}{%
\textit{``There's a rich set of testing libraries such as Holden Karau's Spark Testing Base which is pretty standard in many companies."(P9, Data Engineer, Food Delivery Service)}
    }
}
\end{center}

Pact is a contract testing tool, which is used for testing the schema registry in stream processing messages; ensuring that stream services can communicate with each other.

\begin{center}\noindent\fbox{%
    \parbox{0.6\columnwidth}{%
\textit{``There is the tool to test communication via messages like Pact, it checks when you break the contract of another team."}(P8, Software Engineer, Software Consulting)
    }
}
\end{center}

In the context of fault tolerance, the Chaos Monkey tool was cited. It submits applications to random infrastructure failures. However, this tool does not simulate specific stream failures like message semantics problems.

\begin{center}\noindent\fbox{%
    \parbox{0.6\columnwidth}{%
\textit{``To test fault tolerance we use Chaos Monkey. It creates a chaos scenario in the infrastructure, turns off machines and services, increases latency, affects processing capacity. It tries to force the entire architecture to fail."}(P9, Data Engineer, Food Delivery Service)
    }
}
\end{center}

Participant Q-38 mentions that the WSO2 stream processing tools have features related to stream simulation, data replay and historical and application debugging.

\begin{center}\noindent\fbox{%
    \parbox{0.6\columnwidth}{%
\textit{``WSO2 Stream Processor offers debugging and allows you to replay historical data, simulate random events and replay events in an accelerated way."}(Q-38)
    }
}
\end{center}

There are also reports about in-house tools for system tests. An example is the Ducktape tool cited by Q-1 participant.

\begin{center}\noindent\fbox{%
    \parbox{0.6\columnwidth}{%
\textit{``System testing is done via Ducktape (in-house built tool, but open source). It includes applications correctness test, performance tests, and fault-tolerance tests."} (Q-1)
    }
}
\end{center}


Finally, although many participants claim to build and use their solutions for generating test data. The libraries Clojure Test Check and ScalaCheck were mentioned as a useful tool to build generative tests based on properties. 

\begin{center}\noindent\fbox{%
    \parbox{0.6\columnwidth}{%
\textit{``Clojure has good tools for generative testing, streaming has message contracts, and the applications have entry and exit contracts, so we usually use generative test tools that rely on contracts to generate test data in an automated way."}(P4, Software Engineer, Financial)
}
}
\end{center}


\subsection{Category: Test Data}
\label{sec:cattestdata}

Test data is a subject that emerged a few times in the interviews and questionnaires, with many test data sources being reported, such as replay of historical data, real-time production data mirroring, and synthetic data generators. The quote below exemplifies the use of various test data sources:

\begin{center}\noindent\fbox{%
    \parbox{0.6\columnwidth}{%
\textit{``Replay historical data, real-time use of production data piped to test environment, custom data generation using algorithms. In replay \& real-time data use, the data is often remapped/sanitized for testing \& proper simulation in a test environment."}(Q-100)
    }
}
\end{center}

Although widely adopted, replaying historical data is a sensitive point, especially in contexts where data is confidential, and privacy is a priority, such as banking and financial applications. Therefore, data generation may be the only option for many situations.  The interviewee P2 has expressed concern about privacy with the following comment:

\begin{center}\noindent\fbox{%
    \parbox{0.6\columnwidth}{%
\textit{``The ideal would be the use of historical data, the problem is the restriction of secrecy and privacy, so we work with mock data. We face this barrier in the application of insurance analysis, bank information and restricted information on people."}(P2, Software Architect, Software Consulting)
    }
}
\end{center}

Meanwhile, 8 questionnaire participants mentioned privacy concerns in using historical data, and they suggest data generators as an alternative way of getting around this problem. The following excerpt of Q-96 illustrates this concern from one of the participants in the questionnaire:

\begin{center}\noindent\fbox{%
    \parbox{0.6\columnwidth}{%
\textit{``Mostly replay of historical data. Privacy laws make this less straightforward. We are experimenting with data generators that take into account dependencies between different tables/topics."} (Q-96)
    }
}
\end{center}

Regarding data generators, 59 participants in the questionnaire indicate the use of test data generators. Interviewee P3 mentioned the use of Domain-Specific Languages (DSL) based tools as a technical solution for generating test data. 

\begin{center}\noindent\fbox{%
    \parbox{0.6\columnwidth}{%
\textit{``A tool in which I can easily produce a data in a pattern, for example, a DSL in that it would be effortless to describe the data and that I want to make it happen in the stream."}(P3, Software Engineer, Location Data Services)
    }
}
\end{center}
Although it is a viable alternative in some cases, a random data generator may not generate data that is meaningful to the target application. Participant P11 addressed the importance of using realistic data for testing DSP applications: 

\begin{center}\noindent\fbox{%
    \parbox{0.6\columnwidth}{%
\textit{``We have difficulty in testing fraud detection. We can not always simulate every case. We usually have to simulate forcing some situation of artificial fraud with a fake GPS, which does not represent the scenario of fraud that really happens."} (P11, Data Engineer, Data Science Solutions)
    }
}
\end{center}
In this case, a more robust approach based on artificial intelligence techniques is needed. Thus generators could be intelligent to the point of receiving as input a stream set and be able to detect the format of the messages, standard values and sequential pattern of values on different topics, thus generating multiple test streams. In the questionnaires, three answers reinforce the idea of the need to generate meaningful test data, with one of them quoted next:

\begin{center}\noindent\fbox{%
    \parbox{0.6\columnwidth}{%
\textit{``Generation of meaningful data (not just random) in case there are aggregations or joins within a window, that would allow providing meaningful deterministic output."}(Q-23)
    }
    
}
\end{center}

\section{Results}
\label{sec:discussions}

The previous section presented each category, where participants reported difficulties and relevant aspects when testing DPS applications. The remainder of this section proposes some answers to our research questions based on relationships between categories and literature. 

\textbf{RQ1:} \textit{ What are the approaches and techniques being adopted to test data stream processing applications? }

The traditional methodologies and techniques of software testing remain valid for the context of DSP applications. So, considering data stream particularities, unit tests, integration tests, and system tests are recommended. The comment from P6 in Section~\ref{sec:approaches} motivates this aspect. 

Unit tests are widely employed, and no particular difficulties were reported on unit tests. Participant P7 (Section~\ref{sec:approaches}) reinforces this statement. In the same way, integration testing is being performed with data mocking tools. However, many challenges emerged on system-level tests. The interviewee P2 (Section~\ref{sec:approaches}) addresses this issue.

Due to non-determinism in distributed systems, DSP applications are subject to inconsistencies such as glitches~\cite{cooper2006embedding}. The problems of DSP are revealed when different modules have to interact. Therefore, solutions to support testing of DSP application have to focus on integration and system testing. 

Finally, stream simulation emerged as a specific data stream approach for more realistic system-level testing. Simulation tools exercise specific data stream problems through load tests and fault tolerance simulation. In Section~\ref{sec:approaches}, we present some benefits of using simulators according to practitioners' comments.

\textbf{RQ2:} \textit{ Which test frameworks and tools have been adopted?}

Participants revealed which tools they are using for testing DSP application as well as what are their purposes. Table~\ref{tab:tools} summarizes the most cited tools. In general, practitioners reports in-house built tools, test libraries for stream processing tools, and customisation of standard testing tools. More detailed information is in Section~\ref{sec:cattools}.

\begin{table}[]
\centering
\caption{Most significant adopted tools}
\begin{tabular}{ll}
\hline
\textbf{Purposes Uses}                                                                           & \textbf{Tool Name}                                                                                                  \\ \hline
Contract Testing Tool                                                                            & Pact                                                                                                                \\ \hline
Fault Tolerance                                                                                  & Chaos Monkey                                                                                                        \\ \hline
Integration Tests                                                                                & Local Stack, Jenkins, Flink Test Util
                \\ \hline
Property-based Test                                                                              & ScalaCheck                                                                                                          \\ \hline
\begin{tabular}[c]{@{}l@{}}Stream Simulation,  Data Replay\\ and Data Generation\end{tabular} & WSO2 Stream  Process                                                                                               \\ \hline
\multirow{2}{*}{System Tests}                                                                    & \multirow{2}{*}{\begin{tabular}[c]{@{}l@{}}Pepper-Box plug-in on Jmeter\\ Ducktape\end{tabular}}                    \\
                                                                                                 &                                                                                                                     \\ \hline
\multirow{2}{*}{Unit Test}                                                                       & \multirow{2}{*}{\begin{tabular}[c]{@{}l@{}}Spark  Unit Test  Modules, Mocked \\Streams, Flink Spector\end{tabular}} \\
                                                                                                 
                                                                                                 &                                                                                                                     \\ \hline
\end{tabular}
\label{tab:tools}
\vspace{-6mm}%
\end{table}

\textbf{RQ3:} \textit{ What kind of data is being used to test data stream processing applications?}

Test data stands out as a source of difficulties. The category \textit{Test Data} revealed several test data sources (e.g.,  historical or custom data, mirroring production streams and generators).

Historical data seems to be a good option, and it has been adopted by many companies. The main issue to adopt this technique is the fact that in many cases stream data is sensitive (for instance, financial data). Some companies mitigate this issue by anonymising all historical data. But there are contexts where privacy laws, such as \textit{EU General Data protection regulations}~\cite{voigt2017eu}, significantly restrict access to data. Comments from participants Q-100 and P2 (Section~\ref{sec:cattestdata}) illustrate this issue.

When real or historical data is not available, an alternative is to build a test database or use data generators, such as \textit{Threat Stream Generator} tool~\cite{whiting2008creating}. Participants Q-96 and P3 claims that data generators are being used in their companies.

However, many data generators are not capable of generating consistent test data. Participant P11 presents the importance of realistic data to test fraud detection features. For that matter, the participant Q-23 explains the need to generate significant test data. Therefore, data generators are still an open issue for DSP application. Solutions in that sense can explore statistical models and machine learning techniques to generate more realistic data.

\section{Threats to Validity}
\label{sec:threat}

To obtain willing participants was essential to guarantee that any sensitive data has to be kept confidential to researchers, under human ethics guidelines governing this study. That way, all data were analysed anonymously.

We used a triangulation technique to increase confidence and reduce the  subjectivity of our results~\cite{jonsen2009using}. In our case, we combined interviews (Section~\ref{sec:inter}) and questionnaires (Section~\ref{sec:quest}). Moreover, all categories and themes that emerged from data analysis were extracted and combined by two researchers (the first and second authors). In case of disagreement, the third author was considered as an Oracle.

Some degree of research bias is a common threat related in qualitative research studies. Indeed, other researchers might have different interpretations, and so diverse conclusions may emerge from the same data. The resulting answers, for instance, might be different in other contexts~\cite{denzin2007grounded}. To mitigate this threat, we made available all the material that was produced~\cite{PaperAttachments}. So, any researcher can replicate our analysis or audit our data.

Finally, our study posits itself from the point of view of practitioners with different backgrounds, working in companies in different domains (Section~\ref{sec:demo}). However, we do not claim that our results are valid for other scenarios. 

\section{Conclusion and Future Works}
\label{sec:conclu}

This paper presented an exploratory study on testing DSP applications based on the knowledge and experiences of practitioners. We interviewed 12 professionals, and we also applied questionnaires to other 101 professionals. After that, we adopted a qualitative analysis to find answers to the research questions.

The contribution of this work is to reduce the gap between industry and academia on the knowledge about testing in DSP applications. We explored testing approaches for DSP applications, testing tools, and test data sources. Besides that, we enumerated the main testing tools for data stream applications.


In the future, we plan to evolve this work by carrying out a complete qualitative study based on grounded theory with all data collected. The future research will also be more comprehensive since, in this paper, we concentrated on analysing 1 of 3 sections of interviews and questionnaires, that is, testing DSP applications. Moreover, the next steps in this research can result in a significant source of information about DSP industry practices. In the long term, it could be a first landmark for research topics in DSP. 



\bibliographystyle{unsrt}

\bibliography{references}

\end{document}